\documentstyle[12pt]{article}
\newcommand{\journal}[4]{{\em #1~}#2\,(19#3)\,#4;}
\newcommand{\aihp}{\journal {Ann. Inst. Henri Poincar\'e}}

\newcommand{\jmp}{\journal {J. Math. Phys.}}

\newcommand{\cmp}{\journal {Comm. Math. Phys.}}
\newcommand{\cqg}{\journal {Class. Quantum Grav.}}

\newcommand{\np}{\journal {Nucl. Phys.}}
\newcommand{\pl}{\journal {Phys. Lett.}}

\newcommand{\nc}{\journal {Nuovo Cim.}}

\newcommand{\annp}{\journal {Ann. Phys. (N.Y.)}}

\makeatletter
\@addtoreset{equation}{section}
\makeatother

\def\Lp{\displaystyle{\biggl(}}
\def\Rp{\displaystyle{\biggr)}}

\newcommand{\lp}{\left(}\newcommand{\rp}{\right)}

\renewcommand{\d}{\delta}

\renewcommand{\AA}{{\cal A}}
\newcommand{\BB}{{\cal B}}

\newcommand{\FF}{{\cal F}}
\newcommand{\GG}{{\cal G}}

\newcommand{\LL}{{\cal L}}

\newcommand{\NN}{{\cal N}}

\newcommand{\PP}{{\cal P}}
\newcommand{\QQ}{{\cal Q}}

\newcommand{\VV}{{\cal V}}

\newcommand{\complex}{{\kern .1em {\raise .47ex
\hbox {$\scriptscriptstyle |$}}
    \kern -.4em {\rm C}}}
\newcommand{\real}{{{\rm I} \kern -.19em {\rm R}}}
\newcommand{\rational}{{\kern .1em {\raise .47ex
\hbox{$\scripscriptstyle |$}}
    \kern -.35em {\rm Q}}}
\renewcommand{\natural}{{\vrule height 1.6ex width
05em depth 0ex \kern -.35em {\rm N}}}

\newcommand{\tr}{{\rm {Tr} \,}}

\newcommand{\pa}{\partial}
\newcommand{\pad}[2]{{\frac{\partial #1}{\partial #2}}}

\newcommand{\sla}{\raise.15ex\hbox{$/$}\kern -.57em}

\newcommand{\twiddle}{\lower.9ex\rlap{$\kern -.1em\scriptstyle\sim$}}

\renewcommand{\pad}[2]{{\displaystyle{\frac{\partial #1}{\partial #2}}}}
\renewcommand{\=}{&=&} 
\newcommand{\equ}[1]{(\ref{#1})}

\newcommand{\eq}{\begin{equation}}
\newcommand{\eqn}[1]{\label{#1}\end{equation}}
\newcommand{\eea}{\end{eqnarray}}
\newcommand{\eqa}{\begin{eqnarray}}
\newcommand{\eqan}[1]{\label{#1}\end{eqnarray}}
\newcommand{\ba}{\begin{array}}
\newcommand{\ea}{\end{array}}
\newcommand{\eqac}{\begin{equation}\begin{array}{rcl}}
\newcommand{\eqacn}[1]{\end{array}\label{#1}\end{equation}}

\begin{document}
{\large     


{\ }

\vspace{20mm}
\centerline{\LARGE Algebraic structure of Lorentz and }  \vspace{2mm}

\centerline{\LARGE diffeomorphism anomalies }  \vspace{2mm}

\vspace{9mm}

\centerline{M. Werneck de Oliveira }
\centerline{{\small  International School for Advanced Studies} }
\centerline{{\small  Via Beirut 2-4, Trieste 34014, Italy }}
\vspace{4mm}

\centerline{S.P. Sorella$^{1}$\footnotetext[1]{Supported in part
by the ''Fonds zur F\"orderung der Wissenschaftlichen Forschung'',
M008-Lise Meitner Fellowship.}  }
\centerline{{\small  Institut f\"ur Theoretische Physik} }
\centerline{{\small  Technische Universit\"at Wien }}
\centerline{{\small  Wiedner Hauptstra\ss e 8-10}}
\centerline{{\small  A-1040 Wien (Austria)}}
\vspace{4mm}
\vspace{10mm}

\centerline{{\normalsize {\bf REF. SISSA 24/93/EP}} }

\vspace{4mm}
\vspace{10mm}

\centerline{\Large{\bf Abstract}}\vspace{2mm}
\noindent
The Wess-Zumino consistency conditions for Lorentz and diffeomorphism
anomalies are discussed by introducing an operator $\delta$ which allows
to decompose the exterior space-time derivative as a BRS commutator.

\setcounter{page}{0}
\thispagestyle{empty}

\vfill
\pagebreak
\section{Introduction}

Since the work of L. Alvarez-Gaum{\'e} and E. Witten~\cite{witten},
Lorentz and diffeomorphism anomalies have been the object of continuous
investigations and renewed interests. Also if a
fully satisfactory understanding of the gravitational phenomena is still
lacking, many progress have been done toward a better knowledge of the
peculiar features displayed by these anomalies. For instance,
according to the analysis of ref.~\cite{tonin1} all
known diffeomorphism anomalies can be splitted in two groups, called
first and second-family, which, by means of a non-polynomial Bardeen-Zumino
action~\cite{zumino}, can be converted respectively into Weyl and Lorentz
anomalies~\cite{zumino,tonin2,tonin3,stora,baulieu,ginsparg}.
Moreover, as shown in~\cite{witten}, pure Lorentz anomalies can occur only in
$(4n-2)$, $(n \ge 1)$, space-time dimensions.

We emphasize also that a cohomological algebraic set up, known
as the $descent-equations$ technique, has been developped by several
authors~\cite{tonin1,zumino,stora,baulieu,ginsparg,brandt} to characterize the
gravitational anomalies as non-trivial solutions
of the Wess-Zumino consistency conditions~\cite{wess}. Indeed these conditions,
when formulated in terms of the
corresponding $BRS$ nilpotent operator $s$, lead to the characterization
of the cohomology of $s$ modulo $d$, $d$ being the exterior space-time
derivative, in the space of local polynomials in the fields and their
derivatives. One has then to study the solutions of the equation
\eq
    s \AA{\ }+{\ }d\QQ=0   \ ,
\eqn{consist}
for some local polynomial $\QQ$. $\AA$ is said to be non-trivial if
\eq
    \AA \ne s{\hat \AA} + d{\hat \QQ} \ ,
\eqn{trivial}
with ${\hat \AA}$ and ${\hat \QQ}$ local polynomials. In this case the
integral of $\AA$ on space-time, $\int \AA$, yields the anomaly one is looking
for.
Condition \equ{consist}, due to the relations:
\eq
   s^2{\ } ={\ } d^2{\ } ={\ }sd{\ }+{\ }ds{\ }=0 \ .
\eqn{sd}
and to the algebraic Poincar{\'e} lemma~\cite{cotta,brandt} is easily seen
to generate a ladder of equations:
\eq\ba{lcl}
     s\QQ{\ }{\ }{\ }+{\ }d\QQ^1 = 0   \\
    s\QQ^1{\ }{\ }+{\ }d\QQ^2 = 0   \\
{\ }{\ }{\ }{\ }         ...... \\
{\ }{\ }{\ }{\ }         ...... \\
    s\QQ^{k-1}{\ }+{\ }d\QQ^k = 0   \\
    s\QQ^{k} = 0 \ ,
\ea\eqn{tower}
with $\QQ^i$ local polynomials in the fields.
These equations, as it is well known since several years, can be solved by
means of the so-called homotopy
$"\it russian-formula "$~\cite{zumino,stora,baulieu,ginsparg,dviolette}.

More recently a new way of finding non-trivial solutions of the ladder
\equ{tower} has been proposed by one of the authors~\cite{silvio} and
successfully applied to the study of the Yang-Mills cohomology. The method
relies on the introduction of an operator $\delta$ which allows
to decompose the exterior derivative $d$ as a $BRS$ commutator, i.e.:
\eq
           d{\ }={\ }-[{\ }s{\ },{\ }\d{\ }]   \ .
\eqn{decomposition}
It is very easy to show that repeated applications of $\delta$ on the
polynomial $\QQ^k$ which solves the last equation of \equ{tower} give an
explicit expression for the other cocycles
$\QQ^i$ and for the searched anomaly $\AA$. Moreover, as shown in the
case of Yang-Mills, these expressions turn out to be
cohomologically equivalent to that obtained by the $"\it russian-formula "$.

One has to note also that, actually, the decomposition
\equ{decomposition} represents one of the most interesting features of the
topological field theories. In this case the operator $\delta$ is the
generator of the
topological vector supersymmetry and allows for a complete classification of
anomalies and non-trivial observables for both
Schwarz and Witten-type topological models~\cite{schwarz,birmingham}.

The aim of this work is to extend the analysis of ref.~\cite{silvio} to the
gravitational case. In particular we will show that
decomposition \equ{decomposition} can be used as an alternative tool
for an algebraic
characterization of the Lorentz and the associated second-family
diffeomorphism anomalies.

Let us finish this introduction by making some remarks concerning the
field-theory context which will be adopted in the paper.

In what follows we always refer to the gravitational fields, i.e. to the
vielbein $e$, the spin and the Christoffel connections $(\omega,{\ }\Gamma)$,
as external classical fields. The quantum effective action obtained by
integrating out the matter fields reduces then to a one-loop expansion.

The functional space which will be used to study the cohomology of the $BRS$
operator is assumed to be the space of the integrated local polynomials in the
connections, the Lorentz and diffeomorphism ghosts and their space-time
derivatives. This space, also if does not contain
the non-polynomial Bardeen-Zumino actions~\cite{zumino}, includes all known
Lorentz and diffeomorphism anomalies.
For what concerns the latters we limit
ourselves only to the characterization of the second-family type anomalies
which are known to be related with  pure Lorentz
cocycles~\cite{tonin1,tonin2,tonin3,zumino,stora}. This is due to the fact
that, up to our knowledge, it seems rather difficult to establish an
algorithm analogue to the $"\it russian-formula "$ for the Weyl
trace anomalies. In this case, as
discussed by~\cite{bandelloni} and by~\cite{sibold},
the use of the
so-called Weyl representation~\cite{genova} provides an adequate
algebraic set up. See also ref.~\cite{bast} for a quantum mechanical approach
and ref.~\cite{deser} for a classification based on the properties of the
conformal Weyl tensor.

We take here the conventional point of view of a field theory locally
defined~\cite{tonin1}. As shown in~\cite{stora}, global properties
can be taken into account by the introduction of a fixed background
connection.

Finally, as done in~\cite{zumino,ginsparg}, we have adopted the strategy of
discussing Lorentz and diffeomorphism anomalies as two separate cohomology
problems rather than as a unique one. Let us remark, however, that the
expressions we find for the Lorentz cocycles are easily seen to be
invariant under the diffeomorphism transformations, due to the fact that the
action of the latters on a form-polynomial of maximal space-time
degree reduces to a total derivative. The reciprocal property holds also
for pure diffeomorphism cocycles which, being expressed in terms of the
Christoffel connection, do not contain any Lorentz index.

\section{ Pure Lorentz anomalies}

\subsection{Functional idendities}
Pure Lorentz anomalies~\cite{witten}, due to the fact that local framing
rotations are equivalent to $SO(k)$ ($k$ being the dimension of
the euclidean space-time) gauge transformations, can be studied in strict
analogy with the corresponding Yang-Mills case.
To this purpose, let us introduce the local space $\VV$ of
form-polynomials~\cite{dviolette} in the variables
$(\omega,{\ }d\omega,{\ }\theta,{\ }d\theta)$; $\omega$ and $\theta$ being
respectively the one-form spin connection
$\omega = \omega_{\mu}dx^{\mu} $ and the zero-form Lorentz ghost.
The fields $\omega$ and $\theta$ are Lie algebra valued and are
antisymmetric in the Lorentz indices
\eq
     \omega = {1 \over  2} L^{ab}\omega_{ab} \ , \qquad
     \theta = {1 \over  2} L^{ab}\theta_{ab} \ ,
\eqn{omamatrix}
with
\eq
     \omega_{ab}{\ }={\ }-\omega_{ba}   \ , \qquad
     \theta_{ab}{\ }={\ }-\theta_{ba} \ ,
\eqn{omantis}
and $\{ L^{ab} \}$ the hermitian generators of $SO(k)$ in some suitable
representation. The exterior derivative $d$ is defined by
\eq
     d j_p = dx^\mu \pa_\mu j_p \ ,
\eqn{ddef}
for any $p$-form
\eq
 j_p = {1 \over p!}j_{i_1....i_p}dx^{i_1}....dx^{i_p} \ ,
\eqn{pform}
where a wedge product has to be understood.
The local space $\VV(\omega,d\omega,\theta,d\theta)$ is, as usual, equipped
with a grading given by the sum of the form degree and of the ghost number;
$\omega$ and $\theta$ having respectively ghost number zero and one. In
general a $p$-form with ghost number $q$ will be denoted by $j^q_p$.

The $BRS$ transformations of $\omega$ and $\theta$ are:
\eq\ba{lcl}
 s \omega \= d\theta - i\{ \theta, \omega\} \ ,\\
 s \theta \= -i \theta^2 \ ,
\ea\eqn{brsin}
with
\eq
 s^2{\ }= 0  \ .
\eqn{snilp}
Introducing the two-form Riemann tensor
\eq
   R = d\omega - i\omega^2 \ ,
\eqn{riemann}
one has
\eq
 s R = -i [\theta,R] \ ,
\eqn{brsriem}
and
\eq
 d R = -i [R,\omega] \ ,
\eqn{bianchi}
which expresses the Bianchi identity.

In what follows we will restrict ourselves to finding the non-trivial
solutions of the consistency condition \equ{consist}
belonging to the local space $\VV$; i.e. we will assume that the Lorentz
anomalies are form-valued polynomials in
$(\omega,{\ }d\omega,{\ }\theta,{\ }d\theta)$.
This assumption, although if the local space $\VV$ does not explicitely
contain the
vielbein $e = e_\mu dx^\mu$ and the higher order derivatives of
$\omega$ and $\theta$, is supported by the fact that all known Lorentz
anomalies~\cite{witten,tonin1,zumino,stora,baulieu,ginsparg,brandt} can be
expressed in terms of $(\omega,{\ }d\omega,{\ }\theta,{\ }d\theta)$.
It is worth to mention that
actually, as recently proven by Dubois-Violette et al.~\cite{dvu}, the use of
the space of form-valued
polynomials is not a restriction on the generality of the solutions of the
consistency conditions.

To study the cohomology of $s$ and $d$ we use as independent
variables the set $(\omega,{\ }R,{\ }\theta,{\ }\rho = d\theta)$; i.e. we
replace $d\omega$ with $R$ by using the Bianchi identity \equ{bianchi} and
we introduce the variable $\rho=d\theta$ to stress the local nature of the
condition \equ{consist}. On the local space $\VV(\omega,R,\theta,\rho)$ the
$BRS$ operator $s$ and the exterior derivative $d$ can be represented as
ordinary differential operators:
\eq\ba{lcl}
 s \omega \= \rho - i\{ \theta, \omega\} \ ,\\
 s \theta \= -i \theta^2 \ , \\
 s R      \= -i [\theta,R] \ , \\
 s \rho   \= -i [\theta,\rho] \ ,
\ea\eqn{soper}
and
\eq\ba{lcl}
 d \omega \= R + i\omega^2 \ , \\
 d \theta \= \rho \ , \\
 d R      \= -i [R,\omega] \ ,
\ea\eqn{doper}
One easily verifies that $s$ and $d$ are of degree one and satisfy
\eq
   s^2{\ } ={\ } d^2{\ } ={\ }sd{\ }+{\ }ds{\ }=0 \ .
\eqn{sdeq}

\subsection{The $d$ and $s$ cohomologies}
In this section we briefly recall some useful properties concerning the
cohomologies of $d$ and $s$. Let us begin by showing that, on the local
space $\VV$, the exterior derivative has vanishing cohomology.
Following ref.~\cite{silvio} let us introduce the counting operator
$\NN$
\eq
  \NN \varphi{\ }={\ }\varphi \ , \qquad
          \varphi = (\omega,{\ }R,{\ }\theta,{\ }\rho) \ ,
\eqn{filt}
according to which the exterior derivative \equ{doper} decomposes as
\eq
  d{\ }={\ }d^{(0)}{\ }+{\ }d^{(1)} \ ,
\eqn{ddec}
\eq
 [{\ }\NN{\ },{\ }d^{(\nu)}{\ }]{\ }={\ }\nu d^{(\nu)}  , \qquad \nu=0,1  \ ,
\eqn{nndec}
with
\eq\ba{lcl}
 d^{(0)} \omega \= R \ , \\
 d^{(0)} \theta \= \rho \ ,
\ea\eqn{d0}
and
\eq
 d^{(0)} d^{(0)}{\ }={\ }0 \ .
\eqn{nild0}
Expressions \equ{d0} show that $d^{(0)}$ has vanishing cohomology; it then
follows that also $d$ has vanishing cohomology due to the fact that the
cohomology of $d$ is isomorphic to a subspace of the cohomology of
$d^{(0)}$~\cite{dixon}.

For what concerns the cohomology of $s$ we have the following result:

{\ }

{\ }

{\bf $s$-cohomology}~\cite{tonin1,zumino,stora,ginsparg,brandt}

The cohomology of $s$ on $\VV(\omega,R,\theta,\rho)$ is given by
polynomials in the variables $(\theta,R)$ generated by monomials of the form
\eq
    \Lp \tr {\theta^{2m+1} \over {(2m+1)!}} \Rp{\ } \PP_{2p+2}(R)
  \qquad {\ }{\ }      m,p = 1,2,....      \ ,
\eqn{scohom}
where
\eq
\PP_{2p+2}(R){\ }={\ } \tr R^{p+1}{\ }  \ ,
\eqn{invpol}
is the invariant monomial of degree $(2p+2)$.

{\ }

The indices $(m,p)$ in
eq. \equ{scohom} run according to the set of Casimir invariant tensors which
can be obtained from traces in the representation $\{ L^{ab} \}$.
Moreover, as one can immediately see from eq.\equ{invpol}, the
tensorial group structure of the monomial $\PP_{2p+2}(R)$ contains a totally
symmetric irreducible invariant Casimir tensor of rank $(p+1)$.
Such an invariant tensor exists, however, only if $p=(2n-1)$~\cite{bonora}.
It follows then that pure Lorentz anomalies, besides the abelian case of
$SO(2)$, can occur only in $(4n-2)$ space-time dimension~\cite{witten}.

Due to the Bianchi identity \equ{bianchi} the invariant monomial
$\PP_{4n}(R)$ is $d$-closed
\eq
   d\PP_{4n}(R){\ }=0 \ .
\eqn{closed}
The vanishing of the cohomology of $d$ implies then that $\PP_{4n}(R)$ is
$d$-exact:
\eq
  \PP_{4n}(R){\ }={\ }d\QQ^0_{4n-1}  \ .
\eqn{exact}
Equation \equ{exact}, due to properties \equ{sdeq}, generates a tower of
descent equations:
\eq\ba{lcl}
     s\QQ^0_{4n-1}{\ }+{\ }d\QQ^1_{4n-2} = 0   \\
     s\QQ^1_{4n-2}{\ }+{\ }d\QQ^2_{4n-3} = 0   \\
     {\ }{\ }{\ }{\ }    ......                    \\
     {\ }{\ }{\ }{\ }    ......                    \\
     s\QQ^{4n-2}_1{\ }+{\ }d\QQ^{4n-1}_0 = 0   \\
     s\QQ^{4n-1}_0 = 0   \ ,
\ea\eqn{descequ}
where, according to eq.\equ{scohom}, the non-trivial solution of the last
equation in \equ{descequ} corresponding to the irreducible monomial
$\PP_{4n}(R)$ is given by the ghost monomial of degree
$(4n-1)$:
\eq
      \QQ^{4n-1}_0 {\ }={\ }\tr {\theta^{4n-1} \over {(4n-1)!}} \ .
\eqn{omegsolf}
By definition, the cocycles $\QQ^1_{4n-2}$ and $\QQ^0_{4n-1}$ identify
respectively the Lorentz $SO(4n-2)$ anomaly and the Chern-Simons
$SO(4n-1)$ term. Consistently, the form part of the Riemann tensor in
\equ{exact} is allowed to have components in extra dimensions.

\subsection{Characterization of the solution of the descent equations }

To find a solution of the descent equations \equ{descequ} we proceed as
in~\cite{silvio} and try to decompose the exterior derivative $d$ as a
$BRS$ commutator. This is done by introducing the two operators $\delta$
and $\GG$
\eq\ba{lcl}
  \delta \theta \= - \omega \ ,\\
  \delta \rho   \= R -i \omega^2 \ ,
\ea\eqn{deltaoper}
and
\eq\ba{lcl}
  \GG \theta \= - R \ ,\\
  \GG \rho   \=  -i [R,\omega] \ .
\ea\eqn{ggoper}
These operators are respectively of degree zero $(\delta)$ and one $(\GG)$ and
obey the following algebraic relations
\eq
           d{\ }={\ }-[{\ }s{\ },{\ }\d{\ }]   \ ,
\eqn{alg1}
\eq
           [{\ }d{\ },{\ }\d{\ }]{\ }={\ }2\GG   \ ,
\eqn{alg2}
\eq
   \{{\ }d{\ },{\ }\GG{\ }\}{\ }={\ }0{\ }\qquad \ , \qquad
       \GG \GG{\ }={\ }0  \ ,
\eqn{alg3}
\eq
   \{{\ }s{\ },{\ }\GG{\ }\}{\ }={\ }0{\ }\qquad \ , \qquad
       [{\ }\GG{\ },{\ }\d{\ }]={\ }0  \ .
\eqn{alg4}
One sees then that $\d$ and $\GG$ give rise to the same algebraic structure
as the one already found in~\cite{silvio}.
In particular eq.\equ{alg1} shows that the operator $\d$ decomposes
the exterior derivative $d$ as a $BRS$ commutator. This feature, as it has
been discussed in~\cite{silvio}, allows to characterize a solution of the
ladder \equ{descequ} in a way equivalent to that of the homotopy
$"\it russian-formula "$~\cite{zumino,stora,baulieu,ginsparg,dviolette}.

Indeed, repeating step by step the same procedure of ref.~\cite{silvio}, it is
easy to show that a solution of the tower \equ{descequ} is given by
\eq
      \QQ^{4n-1}_0 {\ }={\ }\tr {\theta^{4n-1} \over {(4n-1)!}} \ ,
\eqn{omegsolut1}
\eq
 \QQ_{2p}^{4n-1-2p}{\ }={\ }{\d^{2p} \over (2p)!} \QQ^{4n-1}_0{\ }-{\ }
       \sum_{j=0}^{p-1} {\d^{2j} \over (2j)!}
         \Omega^{4n-1-2p+2j}_{2p-2j} \ ,
\eqn{even}

for the even space-time form sector and

\eq
     \QQ_1^{4n-2}{\ }={\ }\d \QQ^{4n-1}_0  \ ,
\eqn{omega12sol}
\eq
 \QQ_{2p+1}^{4n-2-2p}{\ }={\ }{\d^{2p+1} \over (2p+1)!}  \QQ^{4n-1}_0
   {\ }-{\ }\sum_{j=0}^{p-1} {\d^{2j+1} \over (2j+1)!}
         \Omega^{4n-1-2p+2j}_{2p-2j} \ ,
\eqn{odd}

for the odd sector and $p{\ }={\ }1, 2,.......,(2n-1)$.

The $\Omega$ expressions in eqs.\equ{even}, \equ{odd} are, as
in~\cite{silvio}, solutions of a second tower of descent equations
which originates from the algebraic relations \equ{alg3}-\equ{alg4} and from
the $s$-cohomology \equ{scohom}. They read:
\eq\ba{lcl}
    \GG  \lp \tr {\theta^{4n-1} \over {(4n-1)!}} \rp{\ }=
          {\ }s \Omega_2^{4n-3} \\
    \GG \Omega_2^{4n-3}{\ }{\ }+{\ }{\ }s \Omega_4^{4n-5}{\ }={\ }0 \\
    \GG \Omega_4^{4n-5}{\ }{\ }+{\ }{\ }s \Omega_6^{4n-7}{\ }={\ }0 \\
{\ }{\ }{\ }{\ }......  \\
{\ }{\ }{\ }{\ }......  \\
    \GG \Omega_{4n-4}^3{\ }{\ }+{\ }{\ }s \Omega_{4n-2}^1{\ }={\ }0 \ .
\ea\eqn{gsdescequ}
and
\eq
  \GG \Omega_{4n-2}^1{\ }={\ }({\em const}) \PP_{4n}(R) \ ,
\eqn{gsend}
where $\PP_{4n}(R)$ is the invariant polynomial of eq. \equ{invpol} and the
$({\em const})$ is an easily computed algebraic factor.

Finally, let us recall that eqs.\equ{omegsolut1}-\equ{odd} identify a
class of solutions which, usually, is not the most general one. However,
as it is well known~\cite{dviolette}, once a particular solution has been
found, the search of the most general solution reduces essentially to a
problem of
$BRS$ local cohomology instead of a modulo-$d$ one. The latter is easily
disentangled by using the
result \equ{scohom} and the algebraic structure
\equ{alg1}-\equ{alg4}.

\subsection{Some examples}
The purpose of this section is to apply the previous construction to
discuss some explicit examples. We will consider, in particular, the cases
of $n=1,2$ which correspond respectively to the two and the six-dimensional
Lorentz anomalies as well as to the three and the seven-dimensional
Chern-Simons terms.

{\bf The case n=1}

In this case, relevant for the $SO(3)$ Chern-Simons term, the descent
equations \equ{descequ} read:
\eq\ba{lcl}
     s\QQ^0_3{\ }+{\ }d\QQ^1_2 = 0   \\
     s\QQ^1_2{\ }+{\ }d\QQ^2_1 = 0   \\
     s\QQ^2_1{\ }+{\ }d\QQ^3_0 = 0   \\
     s\QQ^3_0 = 0   \ ,
\ea\eqn{n1equ}
According to eqs. \equ{omegsolut1}-\equ{odd} and \equ{gsdescequ} the
corresponding solution is given by
\eq
    \QQ^0_3 {\ }={\ }{1 \over 3!} {\d \d \d }\QQ_0^3{\ }-{\ }
                          \d \Omega_2^1   \ ,
\eqn{sl1}
\eq
    \QQ^1_2 {\ }={\ }{1 \over 2} {\d \d  }\QQ_0^3{\ }-{\ }
                           \Omega_2^1   \ ,
\eqn{sl2}
\eq
    \QQ^2_1 {\ }={\ } \d \QQ_0^3  \ ,
\eqn{sl3}
with
\eq
     \QQ^3_0{\ }={\ }\tr {\theta^3 \over 3!} \ ,
\eqn{teta3}
and $\Omega^1_2$ solution of the equation
\eq
  \GG \lp \tr {\theta^3 \over 3!} \rp {\ }={\ }s\Omega^1_2 \ .
\eqn{omega12}
Eq. \equ{omega12} is easily solved with $\Omega^1_2$:
\eq
      \Omega^1_2{\ }={\ }{i \over 2} \tr R\theta  \ .
\eqn{om12sol}
Equations \equ{sl1}-\equ{sl3} become then:
\eq
    \QQ^0_3 {\ }={\ }
         {1  \over 2} \tr \lp  i R\omega - {\omega^3 \over 3} \rp \ ,
\eqn{c3sol10}
\eq
    \QQ^1_2 {\ }={\ }{1 \over 2}\tr \lp \omega \omega \theta -  i R\theta \rp
    {\ }={\ }-{i \over 2} \tr \theta d\omega      \ ,
\eqn{c3sol9}
\eq
    \QQ^2_1 {\ }={\ }-{1 \over 2}\tr \omega \theta \theta \ .
\eqn{c3sol8}
In particular eq.\equ{c3sol10} gives the familiar $SO(3)$ Chern-Simons form.
It is interesting to note that, in spite of the fact that
$\tr {\theta^3 \over 3}$ is identically zero for the Lorentz group $SO(2)$,
the cocycle $\QQ^1_2$ in eq.\equ{c3sol9}, when referred to $SO(2)$,
identifies also the two-dimensional Lorentz anomaly.
In this case, indeed, the $BRS$ transformations \equ{brsin}
become abelian, i.e.:
\eq\ba{lcl}
 s \omega_{12} \= d\theta_{12} \ ,\\
 s \theta_{12} \= 0 \ ,
\ea\eqn{brsab}
$\omega_{12}$ and $\theta_{12}$ being the unique non-zero components of the
spin connection and of the Lorentz ghost. The corresponding cocycle,
$\AA_{abelian}$, is then
\eq
   \AA_{abelian} = \theta_{12} d\omega_{12} \ ,
\eqn{2abanomlay}
and, as shown by explicit computations~\cite{langouche}, turns out to be
the most general expression for the two-dimensional Lorentz anomaly.

{\bf The case n=2}

In this example, relevant for the six-dimensional Lorentz anomaly and for
the seven-dimensional Chern-Simons term, the tower \equ{descequ} takes the
form:
\eq\ba{lcl}
     s\QQ^0_7{\ }+{\ }d\QQ^1_6 = 0   \\
     s\QQ^1_6{\ }+{\ }d\QQ^2_5 = 0   \\
     s\QQ^2_5{\ }+{\ }d\QQ^3_4 = 0   \\
     s\QQ^3_4{\ }+{\ }d\QQ^4_3 = 0   \\
     s\QQ^4_3{\ }+{\ }d\QQ^5_2 = 0   \\
     s\QQ^5_2{\ }+{\ }d\QQ^6_1 = 0   \\
     s\QQ^6_1{\ }+{\ }d\QQ^7_0 = 0   \\
     s\QQ^7_0 = 0   \ .
\ea\eqn{n2equ}
{}From eqs. \equ{omegsolut1}-\equ{odd} a class of solutions is given by
\eq
    \QQ^0_7 {\ }= {\ }{1 \over 7!} {\d}^7 \QQ_0^7{\ }
                    -{\ }{1 \over 5!} {\d}^5 \Omega_2^5{\ }
                    -{\ }{1 \over 3!} {\d}^3 \Omega_4^3{\ }
                    -{\ }{\d} \Omega_6^1  \ ,
\eqn{sol27}
\eq
    \QQ^1_6 {\ }= {\ }{1 \over 6!} {\d}^6 \QQ_0^7{\ }
                    -{\ }{1 \over 4!} {\d}^4 \Omega_2^5{\ }
                    -{\ }{1 \over 2!} {\d}^2 \Omega_4^3{\ }
                    -{\ } \Omega_6^1  \ ,
\eqn{sol26}
\eq
    \QQ^2_5 {\ }= {\ }{1 \over 5!} {\d}^5 \QQ_0^7{\ }
                    -{\ }{1 \over 3!} {\d}^3 \Omega_2^5{\ }
                    -{\ } {\d} \Omega_4^3{\ }  \ ,
\eqn{sol25}
\eq
    \QQ^3_4 {\ }= {\ }{1 \over 4!} {\d}^4 \QQ_0^7{\ }
                    -{\ }{1 \over 2!} {\d}^2 \Omega_2^5{\ }
                    -{\ } \Omega_4^3{\ }  \ ,
\eqn{sol24}
\eq
    \QQ^4_3 {\ }= {\ }{1 \over 3!} {\d}^3 \QQ_0^7{\ }
                    -{\ } {\d} \Omega_2^5{\ } \ ,
\eqn{sol23}
\eq
    \QQ^5_2 {\ }= {\ }{1 \over 2!} {\d}^2 \QQ_0^7{\ }
                    -{\ } \Omega_2^5{\ } \ ,
\eqn{sol22}
\eq
    \QQ^6_1 {\ }= {\ }{\d} \QQ_0^7{\ } \ ,
\eqn{sol21}
and
\eq
    \QQ^7_0{\ }= {\ }\tr {\theta^7 \over 7!} \ .
\eqn{teta7}
The $\Omega$-cocycles are, according to eqs. \equ{gsdescequ}, solutions of
the ladder
\eq\ba{lcl}
    \GG  \lp \tr {\theta^7 \over 7!} \rp{\ }= {\ }s \Omega_2^5  \\
    \GG \Omega_2^5{\ }{\ }+{\ }{\ }s \Omega_4^3{\ }={\ }0 \\
    \GG \Omega_4^3{\ }{\ }+{\ }{\ }s \Omega_6^1{\ }={\ }0 \ ,
\ea\eqn{gsdesn2}
and read
\eq
   \Omega_2^5{\ }={\ }{i \over 6!} \tr \lp R\theta^5 \rp \ ,
\eqn{Omega25}
\eq
   \Omega_4^3{\ }={\ }{1 \over 6!}
            \tr \lp 2R^2\theta^3 + R\theta R\theta^2 \rp \ ,
\eqn{Omega43}
\eq
   \Omega_6^1{\ }={\ }-i{5 \over 6!}
            \tr \lp R^3\theta \rp \ .
\eqn{Omega43}
$\QQ_7^0$ and $\QQ_6^1$ are computed to be
\eq
  \QQ_7^0{\ }={\ }-{5 \over 6!} \tr
  \lp iR^3\omega - {2 \over 5}R^2\omega^3 - {1 \over 5}R\omega R\omega^2
      - {i \over 5}R\omega^5 + {1 \over 35}\omega^7 \rp \ ,
\eqn{cherns7}
\eq\ba{rl}
     \QQ_6^1{\ }= &\!\! {\ }{5 \over 6!}{\ } \tr
   ({\ }  iR^3\theta
   - {2 \over 5}( R^2\omega^2 \theta + R^2\omega\theta\omega
                 +R^2\theta\omega^2 ){\ }) \\
   &\!\! - {1 \over 5} \tr
    ({\ } R\omega R\omega\theta +R\omega R\theta\omega
                         + R\theta R \omega^2{\ } )  {\ }
         + {\ }{1 \over 5} \tr \omega^6 \theta \\
   &\!\! - {i \over 5}\tr
    ({\ } R\omega^4 \theta + R\omega^3 \theta\omega
                       +R\omega^2\theta\omega^2 + R\omega\theta\omega^3
                       +R\theta\omega^4{\ })  \ ,
\ea\eqn{lan6}
and give respectively the seven-dimensional Chern-Simons form and the
six-dimensional Lorentz anomaly. One has to note that, in this case,
expressions \equ{cherns7}-\equ{lan6} do not identify the most general
solution of the descent equations \equ{n2equ}. This is due to the fact that
the local cohomology of $s$ is non-vanishing in the sector with ghost number
three and form-degree four. Indeed from eq.\equ{scohom} it follows that
the factorized monomial
\eq
    \BB^3_4 = \lp \tr R^2 \rp \lp \tr {\theta^3 \over 3} \rp   \ ,
\eqn{bb43}
belongs to the local cohomology of $s$ and, when inserted in the ladder
\equ{n2equ}, gives origin to the sub-tower:
\eq\ba{lcl}
     s\BB^0_7{\ }+{\ }d\BB^1_6 = 0   \\
     s\BB^1_6{\ }+{\ }d\BB^2_5 = 0   \\
     s\BB^2_5{\ }+{\ }d\BB^3_4 = 0   \\
     s\BB^3_4 = 0   \ .
\ea\eqn{n2bbequ}
These equations, thanks to the fact that $\BB^3_4$ is a factorized cocycle, are
easily solved by using the previous results for the case $n=1$
(eqs. \equ{c3sol10} - \equ{c3sol8}). They yield the following solution:
\eq
    \BB^0_7 {\ }={\ }{1  \over 2} \lp \tr R^2 \rp
\lp \tr ( i R\omega - {\omega^3 \over 3})  \rp \ ,
\eqn{bb07}
\eq
    \BB^1_6 {\ }={\ }{1 \over 2} \lp \tr R^2 \rp
\lp \tr ( \omega \omega \theta -  i R\theta ) \rp   \ ,
\eqn{bb16}
\eq
    \BB^2_5 {\ }={\ }-{1 \over 2} \lp \tr R^2 \rp
 \lp \tr \omega \theta \theta  \rp  \ .
\eqn{bb25}
One sees that $\BB^1_6$ and $\BB^0_7$ are given respectively by the
factorized product of $\tr R^2$ with the two-form Lorentz anomaly and the
three-form Chern-Simons term; the only difference being that the trace is
now taken over a representation of $SO(6)$ (resp. $SO(7)$) instead of $SO(2)$
(resp. $SO(3)$). The most general algebraic non-trivial solution for the
six-dimensional Lorentz anomaly is given then by the sum of the irreducible
element \equ{lan6} and of the factorized term in eq.\equ{bb16}. This example
illustrates in a clear way a general feature of the descent equations
\equ{descequ}. It is easy to check indeed that the most general solution of
the ladder \equ{descequ} is usually given by the sum of
an irreducible term (corresponding to the ghost monomial \equ{omegsolf}) and
of the whole set of factorized elements which are solutions of all the
non-trivial
sub-towers allowed by the local cohomology of $s$~\cite{dviolette}.

\section{Pure diffeomorphism anomalies}

\subsection{Generalities}
To characterize the diffeomorphism anomalies let us begin by specifying the
set of variables and the functional space the nilpotent $BRS$ operator acts
upon. Following~\cite{tonin1,zumino,stora,baulieu,ginsparg} we will adopt as
fundamental variable the particular combination of the spin connection
$\omega$ and the vielbein $e$ called Christoffel connection \footnote{ For
simplicity we consider here only the case of vanishing torsion.}:
\eq
         \Gamma = e^{-1} d e + e^{-1}\omega e \ ,
    \qquad \Gamma_{\mu\lambda}^{\rho} = \Gamma_{\lambda\mu}^{\rho} \ .
\eqn{christoffel}
Denoting with $\xi^\mu$ the ghost field associated with the infinitesimal
diffeomorphism transformations, the corresponding $BRS$ operator reads:
\eq\ba{lcl}
 s \Gamma_{\lambda\mu}^{\rho} =
  \xi^{\sigma} \pa_{\sigma} \Gamma_{\lambda\mu}^{\rho} +
  (\pa_{\lambda} \xi^{\sigma}) \Gamma_{\sigma\mu}^{\rho} +
  (\pa_{\mu}     \xi^{\sigma}) \Gamma_{\lambda\sigma}^{\rho} -
  (\pa_{\sigma}    \xi^{\rho}) \Gamma_{\lambda\mu}^{\sigma} -
  \pa_{\lambda}\pa_{\mu}\xi^{\rho}  \ ,  \\
 s \xi^{\mu} = \xi^{\sigma}\pa_{\sigma}\xi^{\mu} \ ,
\ea\eqn{sdiff}
and
\eq
              s^2 = 0 \ .
\eqn{sdiffnilp}
Consequently, the functional space the operator $s$ acts upon is assumed
to be the space $\FF$ of the polynomials in the variables
$(\Gamma, \xi)$ and their space-time derivatives. This
functional local space turns out to be large enough to contain all the
diffeomorphism anomalies; both
those related to the pure Lorentz anomalies and those related to Weyl
anomalies.

For a better understanding of this point let us quote an important result,
valid for any space-time dimension, on the general form of the
diffeomorphism anomalies.

{\ }

{\bf Diffeomorphism anomalies}~\cite{tonin1}

On the space $\FF(\Gamma,\xi)$ the most general
diffeomorphism anomaly $\AA_{diff}$ has the form
\eq
   \AA_{diff} = \int \lp b \pa_\mu \xi^{\mu} +
                      b_{\sigma}^{\mu\nu} \pa_\mu \pa_\nu \xi^{\sigma} \rp \ ,
\eqn{diffanom}
where $b$ is a scalar density and cannot be written as a total derivative
and $b_{\sigma}^{\mu\nu}$ is a tensor under linear $GL$-transformations.
\footnote{ Diffeomorphism transformations \equ{sdiff} reduce to linear
$GL$-transformations in the special case of
$\xi^{\mu}(x)=\alpha^{\mu}_{\sigma}x^{\sigma}$, with $\alpha$ constant
parameters. }

{\ }

The coefficients $b$ and $b_{\sigma}^{\mu\nu}$ define the so-called first
and second-family of diffeomorphism cocycles and
are associated respectively with Weyl and Lorentz
anomalies~\cite{tonin1,zumino,tonin2,tonin3}.

Being interested only in the latter anomalies we focus on the analysis of
the second-family; we proceed then to the algebraic characterization of the
coefficient $b_{\sigma}^{\mu\nu}$. We recall also that, up to the
present time, all known diffeomorphism anomalies of the second-family type
belong to a local space which, as in the Lorentz case, consists
of polynomials of differential
forms~\cite{tonin1,zumino,stora,baulieu,ginsparg,brandt} built with the
Christoffel connection and the Riemann tensor.

We introduce then the space of the form-valued polynomials in the variables
$(\Gamma^{\rho}_{\mu},{\ }R^{\rho}_{\mu},{\ }\Lambda^{\rho}_{\mu},{\ }
\lambda^{\rho}_{\mu})$ where:

$i)$  $\Gamma^{\rho}_{\mu}$ denotes the one-form connection
\eq
        \Gamma^{\rho}_{\mu} = \Gamma_{\lambda\mu}^{\rho} dx^{\lambda} \ ,
\eqn{g-form}
$ii)$  $R^{\rho}_{\mu}$ is the two-form Riemann tensor
\eq
         R^{\rho}_{\mu} = d\Gamma^{\rho}_{\mu}
               - \Gamma^{\rho}_{\sigma} \Gamma^{\sigma}_{\mu} ,
\eqn{rtwoform}
$iii)$ $\Lambda^{\rho}_{\mu}$ is the zero-form
\eq
        \Lambda^{\rho}_{\mu} = -\pa_\mu \xi^{\rho}  \ ,
\eqn{lzeroform}
$iv)$ $\lambda^{\rho}_{\mu}$ is the one-form
\eq
       \lambda^{\rho}_{\mu} =
             {\ }- dx^{\sigma} \pa_{\sigma} \Lambda^{\rho}_{\mu}
             {\ }={\ }-d\Lambda^{\rho}_{\mu}          \ ,
\eqn{loneform}
and $d$ is the ordinary exterior space-time derivative \equ{ddef}.

The diffeomorphism anomalies of the second-family will be defined then as
local polynomials $\AA^{1}_{max}$ in the
variables $(\Gamma,{\ }R,{\ }\Lambda,{\ }\lambda)$ of ghost number
one and maximal space-time form-degree which are non-trivial solutions of the
consistency condition
\eq
    s \AA^{1}_{max} + d \AA^{2}_{max-1} = 0 \ ,
\eqn{diffcond}
where, according to \equ{sdiff}, the action of the $BRS$ operator on the
space of form-polynomials in
$(\Gamma,{\ }R,{\ }\Lambda,{\ }\lambda)$ is defined as

\eq\ba{lcl}
 s \Gamma^{\rho}_{\mu} = \lambda^{\rho}_{\mu} +
           \Gamma^{\rho}_{\sigma}\Lambda^{\sigma}_{\mu} -
           \Gamma^{\sigma}_{\mu}\Lambda^{\rho}_{\sigma} +
           \LL_{\xi}\Gamma^{\rho}_{\mu}   \ ,   \\
 s R     ^{\rho}_{\mu} = \Lambda^{\rho}_{\sigma}R^{\sigma}_{\mu} -
           \Lambda^{\sigma}_{\mu}R^{\rho}_{\sigma} +
           \LL_{\xi}R^{\rho}_{\mu}   \ ,   \\
 s \Lambda^{\rho}_{\mu} = \Lambda^{\rho}_{\sigma}\Lambda^{\sigma}_{\mu} +
              \LL_{\xi}\Lambda^{\rho}_{\mu}   \ ,   \\
 s \lambda^{\rho}_{\mu} = \Lambda^{\rho}_{\sigma}\lambda^{\sigma}_{\mu} -
           \Lambda^{\sigma}_{\mu}\lambda^{\rho}_{\sigma} +
           \LL_{\xi}\lambda^{\rho}_{\mu}   \ .
\ea\eqn{sdiffop}
The symbol $\LL_{\xi}$ in eqs.\equ{sdiffop} denotes the ordinary Lie
derivative, taken with respect to the ghost parameter $\xi$, which acts
only on the form-indices of $(\Gamma,{\ }R,{\ }\Lambda,{\ }\lambda)$ as
for instance:
\eq
  \LL_{\xi} \Gamma^{\rho}_{\mu} =
      \xi^{\sigma}\pa_{\sigma}\Gamma^{\rho}_{\mu} +
      \pa_{\lambda}\xi^{\sigma}\Gamma_{\sigma\mu}dx^{\lambda} \equiv
      \xi^{\sigma}\pa_{\sigma}\Gamma^{\rho}_{\mu} -
      \Lambda^{\sigma}_{\lambda}\Gamma_{\sigma\mu}dx^{\lambda}  \ .
\eqn{lie}
Notice that only one space-time index of the Christoffel
connection \equ{christoffel} has been referred to a form-index.
The $BRS$ transformations \equ{sdiffop} (apart the term
of Lie derivative) have thus the same structure of a gauge transformation
with $\Lambda^{\rho}_{\mu}$ as gauge parameter, $\Gamma^{\rho}_{\mu}$
as gauge connection and with the Riemann tensor $R^{\rho}_{\mu}$
as the ordinary Yang-Mills field strength.

\subsection{Decomposition of the $BRS$ operator}
On the local space $(\Gamma,{\ }R,{\ }\Lambda,{\ }\lambda)$ the $BRS$ operator
of eqs.\equ{sdiffop} naturally decomposes as~\cite{zumino}:
\eq
       s = s_0 + \LL_{\xi} \ ,
\eqn{natdec}
with
\eq\ba{rl}
 s_0 = &\!\!
    ( \lambda^{\rho}_{\mu} + \Gamma^{\rho}_{\sigma}\Lambda^{\sigma}_{\mu}
         - \Gamma^{\sigma}_{\mu}\Lambda^{\rho}_{\sigma} )
          \pad{\ }{\Gamma^{\rho}_{\mu}} +
       \Lambda^{\rho}_{\sigma} \Lambda^{\sigma}_{\mu}
        \pad{\ }{\Lambda^{\rho}_{\mu}} \\
   &\!\!  + ( \Lambda^{\rho}_{\sigma} R^{\sigma}_{\mu} -
          \Lambda^{\sigma}_{\mu} R^{\rho}_{\sigma} )
        \pad{\ }{R^{\rho}_{\mu}} +
        ( \Lambda^{\rho}_{\sigma} \lambda^{\sigma}_{\mu} -
          \Lambda^{\sigma}_{\mu} \lambda^{\rho}_{\sigma} )
        \pad{\ }{\lambda^{\rho}_{\mu}}  \ ,
\ea\eqn{s0}
\eq
         s_0 s_0 = 0 \ ,
\eqn{s0nilp}
One easily
sees that the nilpotent operator $s_0$ in eq.\equ{s0} has exactely the same
form of the corresponding $BRS$-Lorentz operator \equ{soper}; $\Gamma$ and
$\Lambda$ playing the role of the spin-connection $\omega$ and of the Lorentz
ghost $\theta$.

It should be noted also that the action of the Lie derivative
$\LL_{\xi}$ in \equ{natdec} on a form-polynomial of maximal space-time
degree reduces to a total derivative~\cite{zumino}; i.e.
\eq
  \LL_{\xi} \Omega_{max}^{p} = d \Omega_{max-1}^{p+1} \ ,
\eqn{totalderiv}
$\Omega_{max}^{p}$ denoting a form-polynomial of maximal degree and $p$ ghost
number. Now, according to our previous definition, the diffeomorphism anomalies
of the second-family type $\AA^{1}_{max}$ are form-polynomials of maximal
space-time degree which are non-trivial solutions of the consistency
condition \equ{diffcond}. It follows then:
\eq
  \LL_{\xi} \AA^{1}_{max} = d \BB^{2}_{max-1} \ ,
\eqn{tderivanom}
for some local form-polynomial ${\hat \BB}^{2}_{max-1}$.
Equation \equ{tderivanom} tells us that, on the space of forms of maximal
degree, the cohomology of $s$ modulo $d$ and that of $s_0$ modulo $d$ are in
one to one correspondence; the difference between the two operators being a
total derivative. We can thus replace $s$ by $s_0$ in the consistency
equation \equ{diffcond} and characterize $\AA^{1}_{max}$ as a non-trivial
element of the cohomology of $s_0$ modulo $d$.

Moreover, to identify the cohomolgy os $s_0$ modulo $d$, we can use the same
algebraic procedure of the corresponding Lorentz case.
Indeed, one checks that the following algebraic relations hold:
\eq
           d{\ }={\ }-[{\ }s_0{\ },{\ }\d_0{\ }]   \ ,
\eqn{dalg1}
\eq
           [{\ }d{\ },{\ }\d_0{\ }]{\ }={\ }2\GG_0   \ ,
\eqn{dalg2}
\eq
   \{{\ }d{\ },{\ }\GG_0{\ }\}{\ }={\ }0{\ }\qquad \ , \qquad
       \GG_0 \GG_0{\ }={\ }0  \ ,
\eqn{dalg3}
\eq
   \{{\ }s_0{\ },{\ }\GG_0{\ }\}{\ }={\ }0{\ }\qquad \ , \qquad
       [{\ }\GG_0{\ },{\ }\d_0{\ }]={\ }0  \ ,
\eqn{dalg4}
where, in analogy with expressions \equ{doper}, \equ{deltaoper} and
\equ{ggoper}, the operators $d$, $\delta_0$ and $\GG_0$ are given by
\eq
 d = ( R^{\rho}_{\mu} + \Gamma^{\rho}_{\sigma} \Gamma^{\sigma}_{\mu} )
     \pad{\ }{\Gamma^{\rho}_{\mu}} -
     \lambda^{\rho}_{\mu} \pad{\ }{\Lambda^{\rho}_{\mu}} +
     ( \Gamma^{\rho}_{\sigma} R^{\sigma}_{\mu} -
       \Gamma^{\sigma}_{\mu} R^{\rho}_{\sigma} ) \pad{\ }{R^{\rho}_{\mu}} \ ,
\eqn{ddiff}
\eq
 \delta_0 = \Gamma^{\rho}_{\mu} \pad{\ }{\Lambda^{\rho}_{\mu}} +
       ( R^{\rho}_{\mu} - \Gamma^{\rho}_{\sigma} \Gamma^{\sigma}_{\mu} )
       \pad{\ }{\lambda^{\rho}_{\mu}}  \ ,
\eqn{delta0}
and
\eq
 \GG_0 = R^{\rho}_{\mu} \pad{\ }{\Lambda^{\rho}_{\mu}} +
       ( \Gamma^{\rho}_{\sigma} R^{\sigma}_{\mu} -
     \Gamma^{\sigma}_{\mu} R^{\rho}_{\sigma} )
       \pad{\ }{\lambda^{\rho}_{\mu}} \ .
\eqn{gg0}
Expressions \equ{s0}-\equ{s0nilp} and \equ{ddiff} allow to easily
adapt the results
\equ{ddec}-\equ{nild0} and
\equ{scohom} on the Lorentz cohomology to the case of the nilpotent operator
$s_{0}$.
In particular, the exterior derivative \equ{ddiff} turns out to have
vanishing cohomology.

It becomes apparent then that starting with a Lorentz anomaly we can
immediately compute the corresponding second-family diffeomorphism anomaly; it
is sufficient to replace the spin connection $\omega$ with the one-form
Christoffel connection $\Gamma$ and the Lorentz ghost $\theta$ with
the variable $\Lambda$. As an example, the two dimensional diffeomorphism
anomaly is, according to expressions \equ{c3sol9}-\equ{2abanomlay}, given by
\eq
        \int \Lambda^{\rho}_{\mu} d\Gamma^{\mu}_{\rho}  \ .
\eqn{2ddiffanom}

As in the Lorentz case, the operator $\delta_{0}$, and $\GG_{0}$ give a simple
procedure for solving the cohomology of $s_{0}$ modulo $d$ and finding the
second-family diffeomorphism anomalies.

\subsection{A four-dimensional example}
For a better understanding of the algebraic relations
\equ{dalg1}-\equ{dalg4} let us conclude this section by discussing the
construction of a non-trivial four-dimensional diffeomorphism cocycle in the
space of local form-polynomials; this example will give
us the possibility of underlining the importance of the general result
\equ{diffanom} and of evidentiating the difference between the Lorentz and the
diffeomorphism transformations.

We emphasize indeed that, as long as one is concerned only with diffeomorphism
cocycles, the result \equ{diffanom}, being valid in any space-time dimension,
does not forbid the existence of non-trivial solutions of the consistency
condition \equ{diffcond} living in a space-time whose dimensions do not
allow for a Lorentz counterpart. This is the case, for
instance, of the four dimensional space-time.

Let us look then for a non-trivial solution of the consistency condition
\eq
    s_{0} \AA^{1}_4 + d \AA^{2}_3 = 0 \ ,
\eqn{diffcond4}
where $s_{0}$ and $d$ are the operators given in eqs.\equ{s0}, \equ{ddiff} and
$\AA^1_4$ and $\AA^2_3$ are form-polynomials. Equation \equ{diffcond4}, due
to the vanishing of the cohomology of $d$, generates the tower
\eq\ba{lcl}
     s_{0}\AA^2_3{\ }+{\ }d\AA^3_2 = 0   \\
     s_{0}\AA^3_2{\ }+{\ }d\AA^4_1 = 0   \\
     s_{0}\AA^4_1{\ }+{\ }d\AA^5_0 = 0   \\
     s_{0}\AA^5_0 = 0   \ .
\ea\eqn{tower4}
According to the local cohomolgy of $s_0$,
$\AA^5_0$ is given by
\eq
 \AA^5_0 = {1 \over 5!} \Lambda^5 \equiv {1 \over 5!}
    \Lambda^{\mu}_{\nu}\Lambda^{\nu}_{\sigma}\Lambda^{\sigma}_{\tau}
    \Lambda^{\tau}_{\rho}\Lambda^{\rho}_{\mu}  \ .
\eqn{aa50}
As in the Lorentz case, a solution
of eqs.\equ{diffcond4}-\equ{tower4} is easily obtained by acting with
the operator $\delta_0$ of eq.\equ{delta0} on the expression \equ{aa50}.
One easily finds:
\eq
    \AA^1_4 {\ }={\ }{1 \over 4!} {\d_0 \d_0 \d_0 \d_0 }\AA^5_0{\ }-{\ }
                     {1 \over 2}  {\d_0 \d_0}\Omega^3_2{\ }-{\ }
                     \Omega^1_4   \ ,
\eqn{aa14}
\eq
    \AA^2_3 {\ }={\ }{1 \over 3!} {\d_0 \d_0 \d_0  }\AA^5_0{\ }-{\ }
                                  {\d_0 }\Omega^3_2{\ } \ ,
\eqn{aa23}
\eq
    \AA^3_2 {\ }={\ }{1 \over 2} {\d_0  \d_0  }\AA^5_0{\ }-{\ }
                                 \Omega^3_2{\ } \ ,
\eqn{aa32}
\eq
    \AA^4_1 {\ }={\ } {\d_0}\AA^5_0{\ }  \ ,
\eqn{aa14}
where, according to the tower \equ{gsdescequ}-\equ{gsend}, $\Omega^3_2$
and $\Omega^1_4$ are solutions of
\eq\ba{lcl}
    \GG_0  \AA^5_0{\ }= {\ }s_0 \Omega^3_2  \\
    \GG_0 \Omega^3_2{\ }{\ }+{\ }{\ }s_0 \Omega^1_4{\ }={\ }0 \ ,
\ea\eqn{gg0d4}
and computed to be:
\eq
  \Omega^3_2{\ }={\ }-{1 \over 4!} R^{\mu}_{\nu}
      \Lambda^{\nu}_{\sigma}\Lambda^{\sigma}_{\rho}\Lambda^{\rho}_{\mu} \ ,
\eqn{om324}
\eq
  \Omega^1_4{\ }={\ }-{2 \over 4!} R^{\mu}_{\nu} R^{\nu}_{\sigma}
                       \Lambda^{\sigma}_{\mu}  \ .
\eqn{om144}
In particular, the cocycle $\AA^1_4$ is given by
\eq\ba{rl} \AA^1_4{\ }&\!\! =  {1 \over 4!}
   ( \Gamma \Gamma \Gamma \Gamma \Lambda + 2 R R \Lambda
     + R \Gamma \Gamma \Lambda + R \Gamma \Lambda \Gamma
     + R \Lambda \Gamma \Gamma ){\ } \\
   &\!\! \equiv - {2 \over 4!} \Lambda{\ }d
    ( \Gamma d \Gamma - {1 \over 2} \Gamma \Gamma \Gamma ) \ ,
\ea\eqn{andiff14}
and coincides, modulo a $d$-coboundary, with that of
ref.~\cite{tonin1,tonin2,tonin3}. Expression \equ{andiff14} is a
non-trivial solution of the consistency
condition \equ{diffcond4} in the space of the form-polynomials which does
not possess a Lorentz counterpart.
This is due to the fact that the analogue of the ghost monomial \equ{aa50}
in the $SO(4)$-Lorentz case identically vanishes due to the antisymmetric
properties of $\theta$ and to the absence of a third order symmetric
invariant Casimir tensor~\cite{bonora}.

Finally, let us recall that expression \equ{andiff14},
also if cohomologically non-trivial in the local space of form-polynomials,
does not forbid the construction
of an effective quantum action which preserves both Lorentz and
diffeomorphism invariances. It is known indeed that, using the
logarithm of the vielbein as the Goldstone boson field~\cite{zumino}, the
cocycle
\equ{andiff14} can be mapped into zero by a non-polynomial
Bardeen-Zumino~\cite{tonin1} action; according to the well
established result that non-trivial anomalies can arise only
in $(4n-2)$ space-time dimensions~\cite{witten}.

{\ }

\noindent{\large{\bf Acknowledgments}}: We are grateful to
L. Bonora and O. Piguet for useful discussions and comments.


{\ }


}    
\end{document}